# Influence of long-range interaction on degeneracy of eigenvalues of connection matrix of d-dimensional Ising system

B.V. Kryzhanovsky and L.B. Litinskii

Center of Optical Neural Technologies, Scientific Research Institute for System Analysis RAS, Moscow, Russia

E-mail: litin@mail.ru

## Abstract

We examine connection matrices of Ising systems with long-rang interaction on $d$-dimensional hypercube lattices of linear dimensions $L$. We express the eigenvectors of these matrices as the Kronecker products of the eigenvectors for the one-dimensional Ising system. The eigenvalues of the connection matrices are polynomials of the $d$-th degree of the eigenvalues for the one-dimensional system. We show that including of the long-range interaction does not remove the degeneracy of the eigenvalues of the connection matrix. We analyze the eigenvalue spectral density in the limit $L \to \infty$. In the case of the continuous spectrum, for $d \leq 2$ we obtain analytical formulas that describe the influence of the long-range interaction on the spectral density and the crucial changes of the spectrum.

## 1. Introduction

The Ising model is widely used in various science areas. Commonly it describes a system of interacting particles in the nodes of hypercube lattices. In the book [1], one can find a classical review of various approaches to the analysis of the Ising systems and the obtained results. Applications of the Ising model to the studies of phase transitions in solids can be found in the book [2]. The monograph [3] describes applications of this model in spin glasses and neural networks. Following the paper [4], there was a series of publications where the authors used the Ising model for training deep neural networks. A collective monograph [5] describes the relations between the Ising model and the problems of binary optimization. Useful references can also be found in [6].

In the present paper, we obtain exact expressions for the eigenvalues and eigenvectors of the Ising connection matrices on hypercube lattices taking into account interactions with an arbitrary number of neighbors. The exact eigenvalues obtained here can be used when calculating the free energy of a spin system [7], in the analysis of the role of long-range hopping in many-body localization for lattice systems of various dimensions (see [8]-[13] and references therein), and in many other applications.

For natural spin systems, the interaction constants are typically determined by the distances between the spins. Then truncating the number of interactions by accounting only for a finite number of neighbors is an approximation, which holds the better the stronger the interaction decays as a function of distance. However, for artificial spin systems with couplers, such as the ones used for quantum annealing (see for example [14]-[20]), the obtained expressions are exact.

In Section 2, we obtain exact results for the eigenvalues and eigenvectors of the Ising connection matrices with discrete spectra. In Section 3, we present the results for a continuous spectrum of the eigenvalues in the limit $L \to \infty$, where $L$ is a linear size of the system. Section 4 contains discussion and conclusions.

## 2. Eigenvalues spectrum

In this Section, we obtain expressions for the eigenvalues and eigenvectors of the connection matrices of the Ising systems on hypercube lattices with an arbitrary long-range interaction and periodic boundary conditions. We, first, examine one-, two- and three-dimensional lattices and then generalize the results to the case of a hypercube.

## 2.1. 1D Ising model

Let us consider a chain of the length $L$ and set the distance between its nodes to be equal to one. For certainty, we suppose that $L$ is an odd number: $L = 2l+1$. Let $\mathbf{J}(k)$ be an $L \times L$ symmetric matrix that describes the interactions only between spins spaced by the distance $k$ ($k = 1, 2, ..., l$). The structure of the matrix $\mathbf{J}(k)$ is as follows. The ones occupy the $k$-th and ($L-k$)-th its diagonals which are parallel to the main diagonal and the other matrix elements are equal to zero. The central row of the matrix $\mathbf{J}(k)$ has the form

$$(0 \quad \cdots \quad 1 \quad \cdots \quad 0 \quad \cdots \quad 1 \quad \cdots \quad 0),$$

where the ones are at the distance $k$ from the center. We obtain all other rows by consequent cyclic shifts of the $(l+1)$-th row: shifting it to the left we obtain the $l$-th row, the right shift gives the $(l+2)$-row, and so on. For example, when $L=7$, the matrices $\mathbf{J}(1)$, $\mathbf{J}(2)$, and $\mathbf{J}(3)$ are

$$\mathbf{J}(1) = \begin{pmatrix} 0 & 1 & 0 & 0 & 0 & 0 & 1 \\ 1 & 0 & 1 & 0 & 0 & 0 & 0 \\ 0 & 1 & 0 & 1 & 0 & 0 & 0 \\ 0 & 0 & 1 & 0 & 1 & 0 & 0 \\ 0 & 0 & 0 & 1 & 0 & 1 & 0 \\ 0 & 0 & 0 & 0 & 1 & 0 & 1 \\ 1 & 0 & 0 & 0 & 0 & 1 & 0 \end{pmatrix}, \quad \mathbf{J}(2) = \begin{pmatrix} 0 & 0 & 1 & 0 & 0 & 1 & 0 \\ 0 & 0 & 0 & 1 & 0 & 0 & 1 \\ 1 & 0 & 0 & 0 & 1 & 0 & 0 \\ 0 & 1 & 0 & 0 & 0 & 1 & 0 \\ 0 & 0 & 1 & 0 & 0 & 0 & 1 \\ 1 & 0 & 0 & 1 & 0 & 0 & 0 \\ 0 & 1 & 0 & 0 & 1 & 0 & 0 \end{pmatrix}, \quad \mathbf{J}(3) = \begin{pmatrix} 0 & 0 & 0 & 1 & 1 & 0 & 0 \\ 0 & 0 & 0 & 0 & 1 & 1 & 0 \\ 0 & 0 & 0 & 0 & 0 & 1 & 1 \\ 1 & 0 & 0 & 0 & 0 & 0 & 1 \\ 1 & 1 & 0 & 0 & 0 & 0 & 0 \\ 0 & 1 & 1 & 0 & 0 & 0 & 0 \\ 0 & 0 & 1 & 1 & 0 & 0 & 0 \end{pmatrix}.$$

All the matrices $\mathbf{J}(k)$ commute and consequently they all have the same set of the eigenvectors $\{\mathbf{f}_\alpha\}_{\alpha=1}^L$. The components of the vectors $\mathbf{f}_\alpha$ are well-known [21]:

$$f_\alpha^{(i)} = \frac{\cos \varphi_{\alpha i} + \sin \varphi_{\alpha i}}{\sqrt{L}}, \quad \varphi_{\alpha i} = \frac{2\pi(\alpha-1)(i-1)}{L}, \quad i = 1, 2, ..., L. \tag{1}$$

Each matrix $\mathbf{J}(k)$ has its own set of the eigenvalues $\{\lambda_\alpha(k)\}_{\alpha=1}^L$:

$$\mathbf{J}(k)\mathbf{f}_\alpha = \lambda_\alpha(k)\mathbf{f}_\alpha, \quad \lambda_\alpha(k) = 2\cos\left(\frac{2\pi k(\alpha-1)}{L}\right), \quad \alpha = 1, 2, ..., L, \quad k = 1, 2, ..., l.$$

Let $w(k)$ be the constant of interaction between spins that are at the distant $k$ from each other, where $k = 1, 2, ..., l$. Then for the one-dimensional lattice, the interaction matrix taking account for an arbitrary long-range interaction has the form:

$$\mathbf{A}_0 = \sum_{k=1}^l w(k)\mathbf{J}(k).$$

The equations (1) define the eigenvectors of this matrix and its eigenvalues, $\mathbf{A}_0 \mathbf{f}_\alpha = \mu_\alpha \mathbf{f}_\alpha$, are

$$\mu_\alpha = \sum_{k=1}^l w(k)\lambda_\alpha(k), \quad \alpha = 1, 2, ..., L. \tag{2}$$

For simplicity and universality, we introduce the notations,

$$\mathbf{J}(0) = \mathbf{I}, \quad \lambda_\alpha(0) = 1, \quad \alpha = 1, 2, ..., L,$$

where $\mathbf{I}$ is an $L \times L$ unit matrix. In addition, we would like to recall the product rules for matrices and vectors that are the Kronecker products. Suppose we have a matrix $\mathbf{M} = \mathbf{M}_1 \otimes \mathbf{M}_2$ and a vector $\mathbf{F} = \mathbf{F}_1 \otimes \mathbf{F}_2$. Then

$$\mathbf{F}^+ \mathbf{M} \mathbf{F} \equiv \mathbf{F}_1^+ \otimes \mathbf{F}_2^+ \; \mathbf{M}_1 \otimes \mathbf{M}_2 \; \mathbf{F}_1 \otimes \mathbf{F}_2 = \mathbf{F}_1^+ \mathbf{M}_1 \mathbf{F}_1 \cdot \mathbf{F}_2^+ \mathbf{M}_2 \mathbf{F}_2. \tag{3}$$

For the following calculations, the rule (3) is very useful.

## 2.2. 2D Ising model

In this Subsection we discuss the 2D Ising model that is a system of spins in the nods of a square lattice. By $w(m,k)$ we denote the constant of interaction between spins that are shifted from each other at a distance $m$ along one of the lattice axis and at a distance $k$ along the other axis. The connection matrix of such a system is an $L^2 \times L^2$ matrix $\mathbf{B}_0$. It is convenient to present this matrix as an $L \times L$ block matrix

$$\mathbf{B}_0 = \begin{pmatrix} \mathbf{A}_0 & \mathbf{A}_1 & \mathbf{A}_2 & ... & \mathbf{A}_l & \mathbf{A}_l & ... & \mathbf{A}_2 & \mathbf{A}_1 \\ \mathbf{A}_1 & \mathbf{A}_0 & \mathbf{A}_1 & ... & \mathbf{A}_{l-1} & \mathbf{A}_l & ... & \mathbf{A}_3 & \mathbf{A}_2 \\ \mathbf{A}_2 & \mathbf{A}_1 & \mathbf{A}_0 & ... & \mathbf{A}_{l-2} & \mathbf{A}_{l-3} & ... & \mathbf{A}_4 & \mathbf{A}_3 \\ ... & ... & ... & ... & ... & ... & ... & ... & ... \\ \mathbf{A}_l & \mathbf{A}_{l-1} & \mathbf{A}_{l-2} & ... & \mathbf{A}_0 & \mathbf{A}_1 & ... & \mathbf{A}_{l-1} & \mathbf{A}_l \\ ... & ... & ... & ... & ... & ... & ... & ... & ... \\ ... & ... & ... & ... & ... & ... & ... & ... & ... \\ \mathbf{A}_2 & \mathbf{A}_3 & \mathbf{A}_4 & ... & \mathbf{A}_{l-1} & \mathbf{A}_{l-2} & ... & \mathbf{A}_0 & \mathbf{A}_1 \\ \mathbf{A}_1 & \mathbf{A}_2 & \mathbf{A}_3 & ... & \mathbf{A}_l & \mathbf{A}_{l-1} & ... & \mathbf{A}_1 & \mathbf{A}_0 \end{pmatrix}, \tag{4}$$

where $L \times L$ matrices $\mathbf{A}_m$ have the form:

$$\mathbf{A}_m = \sum_{k=0}^{l} w(m,k) \mathbf{J}(k), \quad \mathbf{J}(0) \equiv \mathbf{I}, \quad m = 0, 1, ..., l. \tag{5}$$

In Eq. (5) we set $w(0,0) = 0$ and, consequently, the self-interaction is equal to zero. Note, the central block row of the matrix $\mathbf{B}_0$ is

$$\begin{pmatrix} \mathbf{A}_l & \mathbf{A}_{l-1} & ... & \mathbf{A}_1 & \mathbf{A}_0 & \mathbf{A}_1 & ... & \mathbf{A}_{l-1} & \mathbf{A}_l \end{pmatrix};$$

all other block rows we obtain by evident cyclic shifts.

We can treat a two-dimensional spin system as a set of interacting one-dimensional chains (for example, the horizontal ones.) Then the matrix $\mathbf{A}_0$ describes the interactions between the spins of the one horizontal chain and the matrices $\mathbf{A}_m$ ($m \neq 0$) define interactions between the spins from different chains shifted vertically by $m$ nods.

The matrix $\mathbf{B}_0$ is a block Toeplitz matrix with the matrices $\mathbf{A}_0$ on the main diagonal and the matrices $\mathbf{A}_m$ on its $m$-th and $(L-m)$-th diagonals ($m \neq 0$). It is easy to show that the matrix $\mathbf{B}_0$ is

$$\mathbf{B}_0 = \sum_{m=0}^{l} \mathbf{J}(m) \otimes \mathbf{A}_m = \sum_{m=0}^{l} \sum_{k=0}^{l} w(m,k) \cdot \mathbf{J}(m) \otimes \mathbf{J}(k). \tag{6}$$

Since the matrices $\mathbf{A}_m$ commute, we can write the eigenvectors of the matrix $\mathbf{B}_0$ as the Kronecker products of the eigenvectors (1):

$$\mathbf{F}_{\alpha\beta} = \mathbf{f}_\alpha \otimes \mathbf{f}_\beta, \qquad \alpha, \beta = 1, 2, ..., L. \tag{7}$$

This means that we reduce the eigenvalue problem $\mathbf{B}_0 \mathbf{F}_{\alpha\beta} = \mu_{\alpha\beta} \mathbf{F}_{\alpha\beta}$ to calculation of the value $\mu_{\alpha\beta} = \mathbf{F}_{\alpha\beta}^+ \mathbf{B}_0 \mathbf{F}_{\alpha\beta}$. Then substituting the matrix $\mathbf{B}_0$ in the form (6) and the vector $\mathbf{F}_{\alpha\beta}$ in the form (7) with the aid of the identity (3) we obtain

$$\mu_{\alpha\beta} = \sum_{m=0}^{l} \sum_{k=0}^{l} w(m,k) \cdot \lambda_\alpha(m) \cdot \lambda_\beta(k), \tag{8}$$

where $w(0,0) = 0$ and $\lambda_\alpha(0) = \lambda_\beta(0) = 1$. We see that the eigenvalues $\mu_{\alpha\beta}$ are polynomials of the second degree of the eigenvalues $\lambda_\alpha(k)$ calculated for the one-dimensional system.

As example, let us examine a special case of an isotropic interaction only with the nearest neighbors (the interaction constants are $w(0,1) = w(1,0) = 1$) and the next nearest neighbors (the interaction constant is $w(1,1)$). Then the equation (8) takes the form

$$\mu_{\alpha\beta} = \lambda_\alpha(1) + \lambda_\beta(1) + w(1,1) \cdot \lambda_\alpha(1) \lambda_\beta(1). \tag{9}$$

The equation (9) repeats the result obtained previously in [22, 23] where we discussed this special case.

### 2.3. 3D Ising model

Let us discuss the three-dimensional Ising system of interacting spins that are in the nodes of a cubic lattice. By $w(n,m,k)$ we denote the constant of interaction between spins shifted relative to each other by a distance $n$ along one axis, by a distance $m$ along the other axis, and by a distance $k$ along the third axis. In such a system, it is convenient to write the connection $L^3 \times L^3$ matrix $\mathbf{C}_0$ in the block form:

$$\mathbf{C}_0 = \begin{pmatrix} \mathbf{B}_0 & \mathbf{B}_1 & \mathbf{B}_2 & ... & \mathbf{B}_l & \mathbf{B}_l & ... & \mathbf{B}_2 & \mathbf{B}_1 \\ \mathbf{B}_1 & \mathbf{B}_0 & \mathbf{B}_1 & ... & \mathbf{B}_{l-1} & \mathbf{B}_l & ... & \mathbf{B}_3 & \mathbf{B}_2 \\ \mathbf{B}_2 & \mathbf{B}_1 & \mathbf{B}_0 & ... & \mathbf{B}_{l-2} & \mathbf{B}_{l-3} & ... & \mathbf{B}_4 & \mathbf{B}_3 \\ ... & ... & ... & ... & ... & ... & ... & ... & ... \\ \mathbf{B}_l & \mathbf{B}_{l-1} & \mathbf{B}_{l-2} & ... & \mathbf{B}_0 & \mathbf{B}_1 & ... & \mathbf{B}_{l-1} & \mathbf{B}_l \\ ... & ... & ... & ... & ... & ... & ... & ... & ... \\ \mathbf{B}_3 & \mathbf{B}_4 & \mathbf{B}_5 & ... & \mathbf{B}_{l-2} & \mathbf{B}_{l-3} & ... & \mathbf{B}_1 & \mathbf{B}_2 \\ \mathbf{B}_2 & \mathbf{B}_3 & \mathbf{B}_4 & ... & \mathbf{B}_{l-1} & \mathbf{B}_{l-2} & ... & \mathbf{B}_0 & \mathbf{B}_1 \\ \mathbf{B}_1 & \mathbf{B}_2 & \mathbf{B}_3 & ... & \mathbf{B}_l & \mathbf{B}_{l-1} & ... & \mathbf{B}_1 & \mathbf{B}_0 \end{pmatrix}, \quad (10)$$

where $\mathbf{B}_n$ are $L^2 \times L^2$ matrices ($n = 0,1,2...,l$). To obtain the matrices $\mathbf{B}_n$ we have to generate a set of $L \times L$ matrices $\mathbf{A}_m^{(n)}$,

$$\mathbf{A}_m^{(n)} = \sum_{k=0}^{l} w(n,m,k) \mathbf{J}(k), \quad \mathbf{J}(0) \equiv \mathbf{I}, \quad n,m = 0,1,2,...,l. \quad (11)$$

Since there is no self-interaction, we set $w(0,0,0) = 0$. With the aid of the matrices (11) we generate the matrices $\mathbf{B}_n$:

$$\mathbf{B}_n = \sum_{m=0}^{l} \mathbf{J}(m) \otimes \mathbf{A}_m^{(n)} = \sum_{m=0}^{l} \sum_{k=0}^{l} w(n,m,k) \cdot \mathbf{J}(m) \otimes \mathbf{J}(k).$$

By analogy with the two-dimensional system, we can consider the three-dimensional lattice as a set of interacting planar lattices. Then the matrix $\mathbf{B}_0$ describes the interactions of spins belonging to one (let us say, a horizontal) plane; the matrix $\mathbf{B}_n$ ($n \neq 0$) describes the interactions between the spins from two different planes shifted with respect to each other along the vertical axis by $n$ nodes.

As we see, the matrix $\mathbf{C}_0$ has a form of a block Toeplitz matrix with the matrices $\mathbf{B}_0$ at its main diagonal and the matrices $\mathbf{B}_n$ ($n \neq 0$) at its $n$-th and $(L-n)$-th diagonals. Then, we can write the matrix $\mathbf{C}_0$ as

$$\mathbf{C}_0 = \sum_{n=0}^{l} \mathbf{J}(n) \otimes \mathbf{B}_n = \sum_{n=0}^{l} \sum_{m} \sum_{k=0}^{l} w(n,m,k) \cdot \mathbf{J}(n) \otimes \mathbf{J}(m) \otimes \mathbf{J}(k).$$

Since the matrices $\mathbf{B}_n$ commute, we can write the eigenvectors of the matrix $\mathbf{C}_0$ as the Kronecker products of the eigenvectors (1):

$$\mathbf{F}_{\alpha\beta\gamma} = \mathbf{f}_\alpha \otimes \mathbf{f}_\beta \otimes \mathbf{f}_\gamma, \quad \alpha,\beta,\gamma = 1,2,...,L.$$

This means that we reduce the eigenvalues problem $\mathbf{C}_0 \mathbf{F}_{\alpha\beta\gamma} = \mu_{\alpha\beta\gamma} \mathbf{F}_{\alpha\beta\gamma}$ to calculation of the values $\mu_{\alpha\beta\gamma} = \mathbf{F}_{\alpha\beta\gamma}^+ \mathbf{C}_0 \mathbf{F}_{\alpha\beta\gamma}$ and with account for Eq. (3), we obtain:

$$\mu_{\alpha\beta\gamma} = \sum_{n=0}^{l} \sum_{m=0}^{l} \sum_{k=0}^{l} w(n,m,k) \cdot \lambda_\alpha(n) \cdot \lambda_\beta(m) \cdot \lambda_\gamma(k), \quad (12)$$

where, as usually, $\lambda_\alpha(0) = \lambda_\beta(0) = \lambda_\gamma(0) = 1$. We see that in the three-dimensional case the eigenvalues are the polynomials of the third degree of the eigenvalues for the one-dimensional system.

As an example, let us discuss a special case of the three-dimensional isotropic Ising system that is $w(n,m,k) = w(k,n,m) = w(m,n,k)$. We suppose that only the interactions with the nearest neighbors, the next nearest and the third neighbors are nonzero. We set $w(0,0,1) = w(0,1,0) = w(1,0,0) = 1$, $w(0,1,1) = w(1,0,1) = w(1,1,0) = b_1$, and $w(1,1,1) = b_2$. Then from Eq. (12) we obtain

$$\mu_{\alpha\beta\gamma} = w_1 \left( \lambda_\alpha(1) + \lambda_\beta(1) + \lambda_\gamma(1) \right) + w_2 \left( \lambda_\alpha(1) \lambda_\beta(1) + \lambda_\alpha(1) \lambda_\gamma(1) + \lambda_\beta(1) \lambda_\gamma(1) \right) + w_3 \lambda_\alpha(1) \lambda_\beta(1) \lambda_\gamma(1).$$

The last expression coincides with result obtained previously in [22, 23], where we examined the same special case.

### 2.4. Ising system on hypercube

A generalization to the case of a $d$-dimensional lattice is evident. Let us introduce the constants of interaction between spins $w(k_1, k_2, ..., k_d)$, where $k_1$ is a relative distance between the spins along the axis 1, $k_2$ is a relative distance between the spins along the axis 2, and so on. We generate a set of $L \times L$ matrices

$$\mathbf{A}_{k_2}^{(k_3,...,k_d)} = \sum_{k_1=0}^{l} w(k_1, k_2, ..., k_d) \mathbf{J}(k_1),$$

where $k_i = 0, 1, ..., l$, and $i = 2, 3, ..., d$. We use these matrices to generate a set of $L^2 \times L^2$ matrices $\mathbf{B}_{k_3}^{(k_4,...,k_d)} = \sum_{k_2=0}^{l} \mathbf{J}(k_2) \otimes \mathbf{A}_{k_2}^{(k_3,...,k_d)}$. In the same way the matrices $\mathbf{B}_{k_3}^{(k_4,...,k_d)}$ allows us to construct the matrices $\mathbf{C}_{k_4}^{(k_5,...,k_d)} = \sum_{k_3=0}^{l} \mathbf{J}(k_3) \otimes \mathbf{B}_{k_3}^{(k_4,...,k_d)}$ for the tree-dimensional system, and so on until we obtain the matrix $\mathbf{U}_0^{(d)}$ that defines the interactions in the $d$-dimensional system. This is a block matrix similar to the matrices (4) and (10), where the blocks are the $L^{d-1} \times L^{d-1}$ matrices that describe interactions in the $(d-1)$-dimensional system. Omitting the intermediate calculations, we obtain

$$\mathbf{U}_0^{(d)} = \sum_{k_1=0}^{l} \sum_{k_2=0}^{l} ... \sum_{k_d=0}^{l} w(k_1, k_2, ..., k_d) \cdot \mathbf{J}(k_1) \otimes \mathbf{J}(k_2) \otimes ... \otimes \mathbf{J}(k_d), \tag{13}$$

where $w(0, 0, ..., 0) = 0$, and $\mathbf{J}(0) \equiv \mathbf{I}$.

Since in Eq. (13) the matrices $\mathbf{J}(k)$ commute, the eigenvectors of the matrix $\mathbf{U}_0$ are the Kronecker products of the eigenvectors for the one-dimensional system:

$$\mathbf{F}_{\alpha_1 \alpha_2 ... \alpha_d} = \mathbf{f}_{\alpha_1} \otimes \mathbf{f}_{\alpha_2} \otimes ... \otimes \mathbf{f}_{\alpha_d}, \quad \alpha_i = 1, 2, ..., L, \quad i = 1, 2, ..., d.$$

Then again we reduce the eigenvalue problem $\mathbf{U}_0^{(d)} \mathbf{F}_{\alpha_1 \alpha_2 ... \alpha_d} = \mu_{\alpha_1 \alpha_2 ... \alpha_d} \mathbf{F}_{\alpha_1 \alpha_2 ... \alpha_d}$ to calculation of the values $\mu_{\alpha_1 \alpha_2 ... \alpha_d} = \mathbf{F}_{\alpha_1 \alpha_2 ... \alpha_d}^{+} \mathbf{U}_0^{(d)} \mathbf{F}_{\alpha_1 \alpha_2 ... \alpha_d}$. With account for Eq. (3) we obtain

$$\mu_{\alpha_1 \alpha_2 ... \alpha_d} = \sum_{k_1=0}^{l} \sum_{k_2=0}^{l} ... \sum_{k_d=0}^{l} w(k_1, k_2, ..., k_d) \prod_{i=1}^{d} \lambda_{\alpha_i}(k_i). \tag{14}$$

## 3. Density of eigenvalue spectrum

In the previous Sections, we obtained the expressions for the eigenvalues of the connection matrices in multidimensional Ising systems, which allow us to estimate the degeneracy of their spectra. However, in the limit $L \to \infty$ it is more efficient to pass from the discrete to continuous spectrum and analyze the spectral density $P(\mu)$ of the eigenvalue spectrum, where $P(\mu) d\mu$ is the number of the eigenvalues in the interval $[\mu, \mu + d\mu]$. In this limit we succeed in deriving analytical expressions only for one- and two-dimensional systems when we account for interactions with the nearest and the next nearest neighbors. For certainty, we suppose that the constant of interaction with the next nearest neighbor $b$ is positive: $b \geq 0$.

### 3.1. Spectrum density of 1D system

Let us examine the one-dimensional system. Let $w(1) = 1$ and $w(2) \equiv b$ be the constants of interaction with the nearest and the next nearest spins, respectively. Then from Eq. (2) it follows that $\mu_\alpha = 2\cos\varphi_\alpha + 2b\cos 2\varphi_\alpha$, where $\varphi_\alpha = 2\pi(\alpha-1)/L$ and $\alpha = 1, 2, ..., L$. We obtain the density $P(\mu)$ by integrating the delta function $\delta(\mu - 2\cos\varphi_\alpha + 2b\cos 2\varphi_\alpha)$ with respect to the variable $\varphi_\alpha \in [0, 2\pi]$.

i) We begin with the simplest case $b = 0$. After rather simple calculations we obtain

$$P(\mu) = \frac{L}{\pi\sqrt{4-\mu^2}}, \quad -2 < \mu < 2.$$

We see that the spectrum is limited to the interval $\mu \in (-2, 2)$ and there is a divergence at the spectrum ends.

ii) Let examine the values of $b$ inside the interval $0 < b < 1/4$. In this case the spectral density is nonzero only inside the interval $\mu \in (-2+2b, 2+2b)$, where

$$P(\mu) = \frac{L}{2\pi} \frac{Q_0}{\sqrt{(2+2b-\mu)(\mu+2-2b)}}, \quad -2+2b \leq \mu \leq 2+2b. \tag{15}$$

In Eq. (15) $Q_0 = Q_0(\mu)$ is a slow function without singularities on the interval in question:

$$Q_0 = \left| \frac{\left(1 + 4b + \sqrt{1 + 4b(\mu + 2b)}\right)\left(1 - 4b + \sqrt{1 + 4b(\mu + 2b)}\right)}{1 + 4b(\mu + 2b)} \right|^{1/2}.$$

We see that account for the interaction with the next nearest neighbors shifts the spectrum in the positive direction by a value equal to $2b$. In the same time, divergences at the ends of the spectrum ($\mu \to \pm 2 + 2b$) remain. Note, when

$b \to 1/4$ the left end of the spectrum, where $P(\mu) \approx L/\pi\sqrt{(1-4b)(\mu+2-2b)}$, is much higher its right end where $P(\mu) \approx L/\pi\sqrt{(1+4b)(2+2b-\mu)}$.

iii) When $b = 1/4$ the spectral density is nonzero on the interval $\mu \in [-3/2, 5/2]$. In this case Eq. (15) takes the form

$$P(\mu) = \frac{L}{2\pi} \frac{\left(2+\sqrt{3/2+\mu}\right)^{1/2}}{(3/2+\mu)^{3/4}(5/2-\mu)^{1/2}}, \quad -3/2 \leq \mu \leq 5/2.$$

As we see, in this case the divergence at the left end of the spectrum ($\mu \to -3/2$) is much stronger.

iv) Let us analyze the values of $b > 1/4$. We introduce the notations

$$\mu_{min} = -2b - \frac{1}{4b}, \quad \mu_{av} = -2+2b, \quad \mu_{max} = 2+2b.$$

Here $\mu_{min}$ and $\mu_{max}$ are the lower and the upper boundaries of the spectrum, respectively. Next, $\mu_{av}$ is a point between the spectrum boundaries: $\mu_{min} < \mu_{av} < \mu_{max}$. The spectrum is a composite: in the different regions of the interval it is described by two different functions that does not match

$$P(\mu) = \frac{L}{\pi\sqrt{\mu-\mu_{min}}}\left(\frac{Q_1}{\sqrt{\mu_{av}-\mu}} + Q_2\right) \quad \text{when} \quad \mu_{min} \leq \mu < \mu_{av},$$

$$P(\mu) = \frac{L}{\pi Q_1 \sqrt{(\mu-\mu_{min})(\mu_{max}-\mu)}} \quad \text{when} \quad \mu_{av} \leq \mu \leq \mu_{max},$$

where $Q_1 = Q_1(\mu)$ and $Q_2 = Q_2(\mu)$ are slow functions without singularities at the abovementioned intervals

$$Q_1 = \left|\frac{4b-1+\sqrt{4b(\mu-\mu_{min})}}{4b+1+\sqrt{4b(\mu-\mu_{min})}}\right|^{1/2}, \quad Q_2 = \frac{\sqrt{4b}}{Q_1 \left|(4b+1)^2 - 4b(\mu-\mu_{min})\right|^{1/2}}.$$

We see that at the point $\mu = \mu_{av}$ the function $P(\mu)$ is discontinues: when $\mu \to \mu_{av} - 0$, there is a singularity $P(\mu) \sim (\mu-\mu_{min})^{-1/2}$; when $\mu \to \mu_{av} + 0$ the spectrum density is finite: $P(\mu) = 2bL/\pi(4b-1)^{3/2}$. In Fig. 1, we show how the spectrum changes when $b$ increases.

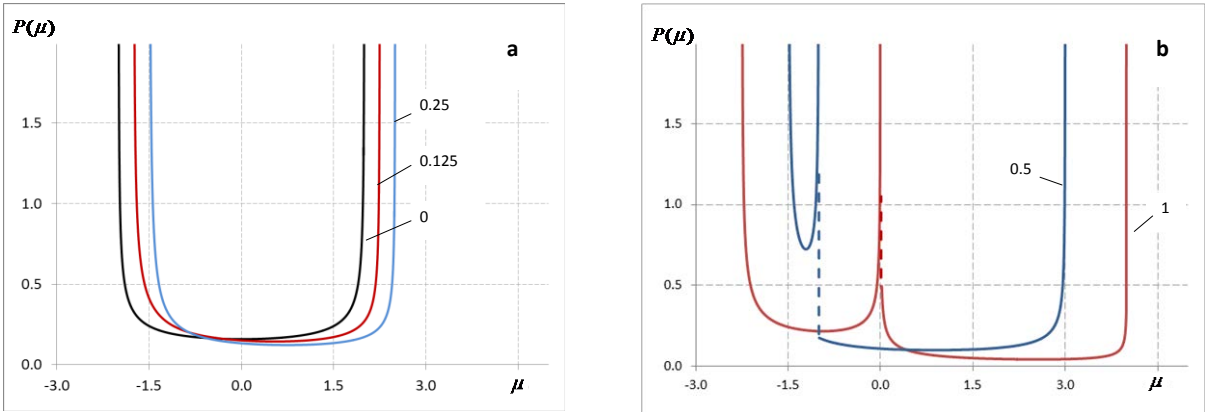

**Fig. 1.** Spectral density $P(\mu)$ for 1D Ising system for different values of parameter $b$. (a) $0 \leq b \leq 0.25$; we show graphs for $b = 0, 0.125, 0.25$. (b) $b > 0.25$; we show graphs for $b = 0.5$ and $1$.

The following picture arises from our analysis. If $b \leq 1/4$, the spectrum as a whole shifts by the value $2b$; the lower and upper spectrum boundaries are $\mu_{min} = -2+2b$ and $\mu_{max} = 2+2b$, respectively; the width of the spectrum remains constant. When $b > 1/4$, an increase of $b$ leads to the spectrum broadening. Namely, the value of $\mu_{min}$ decreases: $\mu_{min} = -2b - 1/4b$ and the value of $\mu_{max}$ increases: $\mu_{max} = 2+2b$. The spectrum width also increases as $\mu_{max} - \mu_{min} = 2+4b+1/4b$. For an arbitrary $b$ there are divergences both at the lower and upper spectrum boundaries. Moreover, if $b > 1/4$ an additional divergence at the point $\mu = \mu_{av} \equiv -2+2b$ appears. In Fig. 2a, we show this picture schematically.

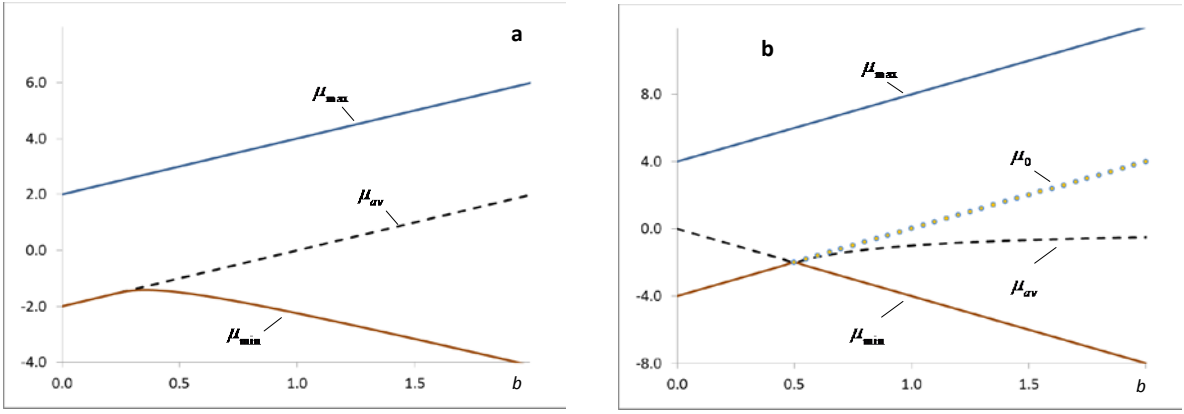

**Fig. 2.** Dependences of values $\mu_{\max}$, $\mu_{\min}$, $\mu_{av}$ on interaction constant $b$. Solid lines correspond to $\mu_{\min}$ and $\mu_{\max}$. (a) 1D Ising model. Dashed line describes movement of singular point $\mu_{av}$, which appears when $b > 1/4$. (b) 2D Ising model. Dashed line and dotted line show movement of singular point $\mu_{av}$ and jump point $\mu_0$, respectively.

### 3.2. Spectral density of 2D system

We analyze the two-dimensional Ising system in the simplest case supposing an isotropic interaction and accounting for interactions with the nearest and next nearest neighbors only. Let the constant of interaction with the nearest neighbors be equal to one and by $b = w(1,1)$ we denote the interaction with the next nearest neighbors. Then from Eq. (9), we obtain $\mu_{\alpha\beta} = 2\cos\varphi_\alpha + 2\cos\varphi_\beta + 4b\cos\varphi_\alpha 2\cos\varphi_\beta$, where $\varphi_\alpha = 2\pi(\alpha-1)/L$, and $\varphi_\beta = 2\pi(\beta-1)/L$. Consequently, to obtain the spectral density $P(\mu)$ we have to integrate the delta function $\delta(\mu - \mu_{\alpha\beta})$ with respect to the variables $\varphi_\alpha \in [0, 2\pi]$ and $\varphi_\beta \in [0, 2\pi]$.

i) We begin with the case $0 \le b \le 1/2$, where the spectral density is nonzero only when $-4 + 4b \le \mu \le 4 + 4b$. When $b = 0$, we obtain

$$P(\mu) = \frac{L^2}{2\pi^2} K(m), \text{ where } m = 1 - \frac{\mu^2}{16}, \ \mu \in [-4, 4]. \tag{16}$$

In Eq. (16), $K(m)$ is a complete elliptic integral of the first kind. We see that the spectrum is symmetric with respect to the point $\mu = 0$ and since $K(m) \approx -\ln|\mu|$ at $m \to 1$, it has a logarithmic divergence when $\mu \to 0$ (see Fig. 3a).

When $b \neq 0$ ($b < 1/2$), we obtain a more general expression

$$P(\mu) = \frac{L^2}{2\pi^2 \sqrt{1+b\mu}} K(m), \ m = \frac{1-(b-\mu/4)^2}{1+b\mu} \text{ when } \mu \in [-4+4b, 4+4b]. \tag{17}$$

As it follows from Eq. (17), when $b$ increases, the spectrum as a whole shifts to the right by the value of $4b$. In the same time, the maximum of the function $P(\mu)$ at $\mu = -4b$, where there is the logarithmic singularity at $m \to 1$, shifts to the left. In addition, at the right end of the spectrum ($\mu = 4 + 4b$) the value of the spectral density decreases as $P(\mu) == L^2/4\pi(1+2b)$ and at the left end ($\mu = -4 + 4b$) it increases as $P(\mu) == L^2/4\pi(1-2b)$. We show the changes of $P(\mu)$ in Fig. 3a.

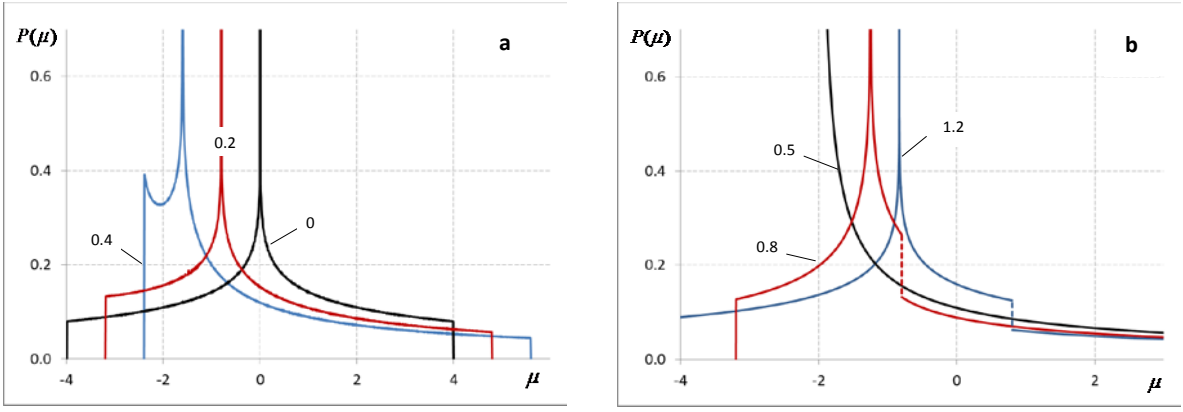

**Fig. 3.** Spectral density $P(\mu)$ for 2D Ising system and different values of constant of interaction with next nearest neighbors. (a) When $0 \leq b < 0.5$, $b = 0,\ 0.2$, and $0.4$; (b) When $b \geq 0.5$, $b = 0.5,\ 0.8$, and $1.2$. Dotted lines define jumps of spectral density.

The spectrum changed drastically when $b$ reaches the value 0.5. In this case, Eq. (16) takes the form
$$P(\mu) = \frac{L^2}{2\pi^2 \sqrt{1 + \mu/2}} K(m),\ m = \frac{6-\mu}{8}\ \text{when}\ \mu \in [-2,\ 6].$$
As we see, the entire spectrum is to the right from the point $\mu = -2$. Near this point when $\mu \to -2$ and, consequently, $m \to 1$, we have $P(\mu) \sim -\ln|\mu+2|/\sqrt{\mu+2}$. Then the density has both the logarithmic and power divergences. When $\mu$ increases the value of $P(\mu)$ decreases gradually and reaches its minimal value $P(\mu) = L^2/8\pi$ at the right end of the spectrum where $\mu = 6$.

ii) The case $b > 1/2$. Now the spectral density is nonzero at the interval $-4b \leq \mu \leq 4 + 4b$. There are three parts of this interval where the discontinuous function $P(\mu)$ is described by three different expressions:
$$P(\mu) = \frac{L^2}{2\pi^2} R\,K(m),$$
where
$$R = \frac{2}{\sqrt{(b - \mu/4)^2 - 1}},\quad m = \frac{(b + \mu/4)^2}{(b - \mu/4)^2 - 1}\quad \text{when}\ -4b \leq \mu < -1/b;$$
$$R = \frac{2}{b + \mu/4},\quad m = \frac{(b - \mu/4)^2 - 1}{(b + \mu/4)^2}\quad \text{when}\ -1/b < \mu \leq -4 + 4b;$$
$$R = \frac{1}{\sqrt{1 + b\mu}},\quad m = \frac{1 - (b - \mu/4)^2}{1 + b\mu}\quad \text{when}\ -4 + 4b \leq \mu \leq 4 + 4b.$$

At the left end of the first interval $\mu \in [-4b, -1/b]$, the density $P(\mu)$ has a finite value $P(\mu) = L^2/2\pi\sqrt{4b^2 - 1}$; at the right end of this interval there is a logarithmic divergence since $m \to 1$ and $P(\mu) \sim -\ln|\mu + 1/b|/(4b^2 - 1)$ when $\mu \to -1/b - 0$. At the left end of the second interval there is also the same logarithmic discontinuity when $\mu \to -1/b + 0$ and at the right end of this interval the function $P(\mu)$ has a finite value $P(\mu) = L^2/2\pi(2b - 1)$. The function $P(\mu)$ has no singularities on the third interval $\mu \in [-4 + 4b,\ 4 + 4b]$. When $\mu$ increases, the density decreases smoothly from the value $P(\mu) = L^2/4\pi(2b - 1)$ when $\mu = -4 + 4b$ up to the value $P(\mu) = L^2/4\pi(2b + 1)$ when $\mu = 4 + 4b$. In Fig. 3 b., we show all the transformations of the function $P(\mu)$ when $b$ increases.

Let us summarize the analysis of this Subsection. When $b$ changes inside the interval $0 \leq b \leq 1/2$, the entire spectrum shifts to the right by the value of $4b$: the lower boundary of the spectrum is $\mu_{\min} = -4 + 4b$ and its upper boundary is $\mu_{\max} = 4 + 4b$. Inside the interval there is a logarithmic discontinuity at the point $\mu = \mu_{av}$ ($\mu_{av} = -4b$ when $0 \leq b \leq 1/2$). When $b > 1/2$, the further increase in $b$ leads to the broadening of the spectrum and change of its form. The lower spectrum boundary $\mu_{\min} = -4b$ decreases and its upper boundary $\mu_{\max} = 4 + 4b$ increases. The spectrum density has two singularities: the divergence at $\mu = \mu_{av}$ and the finite jump at $\mu = \mu_0$, where $\mu_{av} = -1/b$ and $\mu_0 = -4 + 4b$. Fig. 2b presents this picture schematically.

### 3.3. Spectrum density for $d \geq 3$

In the case of Ising systems of higher dimensions, due to significant mathematical difficulties we were unable to obtain analytical expressions for the spectral density. However, it is possible to make some general conclusions. For example, when we account for the nearest neighbors only, we can present the spectral density $P_d(\mu)$ for the $d$-dimensional system as a convolution of the spectral densities of the systems of lower dimensions:

$$P_d(\mu) = \int_{-\infty}^{\infty} P_{d_1}(x) P_{d_2}(\mu - x) dx, \quad d_1 + d_2 = d.$$

Due to the integration the divergence of $P_d(\mu)$ is weaker the divergences of $P_{d_1}(x)$ and $P_{d_2}(x)$. However, there are many cases when the density $P_d(\mu)$ has singularities. For example, suppose that $P_{d_1}(x)$ and $P_{d_2}(x)$ have rather strong divergences at the points $x_1$ and $x_2$, respectively: $P_{d_1}(x \to x_1) \sim |x - x_1|^{-r_1}$ and $P_{d_2}(x \to x_2) \sim |x - x_2|^{-r_2}$, where $0.5 \leq r_1, r_2 < 1$. Then the density $P_d(\mu)$ ($d = d_1 + d_2$) also has a singularity at the point $\mu = x_1 + x_2$. This may be a power singularity $P_d(\mu \to x_1 + x_2) \sim |\mu - x_1 - x_2|^{-r}$ when $r > 0$, or a logarithmic singularity $P_d(\mu \to x_1 + x_2) \sim -\ln|\mu - x_1 - x_2|$ when $r = 0$. In both formulas $r = r_1 + r_2 - 1$. Moreover, the singularity may be a combination of the logarithmic and power singularities. To avoid misunderstanding we note that the functions $P_{d_1}(x)$ and $P_{d_2}(x)$ are normalized functions and according definition may have only integrable singularities.

## 4. Discussion and conclusions

The analysis of the discrete spectra in Section 2 shows that when the dimension of the space $d$ increases the degeneracy of the eigenvalues increases too. Let us discuss the influence of account for the long-range interaction on the spectrum degeneracy.

In the one-dimensional system, all the eigenvalues of the interaction matrix are two-fold degenerate excluding the non-degenerate eigenvalue $\mu_1 = \sum_{k=1}^{l} w(k) \lambda_1(k) = 2\sum_{k=1}^{l} w(k)$. Such degeneracy takes place when we account for the interaction with the nearest neighbors and, as it follows from Eq. (2), it does not change when we include long-range interactions.

The situation becomes more complicated when the dimension of the lattice is larger. Let us start with the isotropic case when the interaction constant $w(k_1, k_2, ..., k_d)$ does not change when rearranging the numbers $k_1, k_2, ..., k_d$. As a result, the eigenvalue $\mu_{\alpha_1 \alpha_2 ... \alpha_d}$ defined by Eq. (14) does not depend on the order of the indices $\alpha_1, \alpha_2, ..., \alpha_d$, where $\alpha_i \in [0, L]$, $i = 1, 2, .., d$. For example, in the two-dimensional system $w(n, m) = w(m, n)$ and $\mu_{\alpha\beta} = \mu_{\beta\alpha}$. In the isotropic $d$-dimensional case, the degeneracy of the eigenvalue $\mu_{\alpha_1 \alpha_2 ... \alpha_d}$ is equal to $2^{d-r} d!$, where $r$ is the number of indices that are equal to one. The factor $2^{d-r}$ appears due to the double degeneracy of the eigenvalues in the one-dimensional model and the factor $d!$ is equal to the number of permutations of the indices $\alpha_1, \alpha_2, ..., \alpha_d$. The same is true if we account only for interaction with the nearest neighbors and the result does not change when we include long-range interactions.

In anisotropic systems, where the interaction constants along different axes are not the same the situation is somewhat different. In this case, the value of $w(k_1, k_2, ..., k_d)$ changes when we rearrange the arguments $k_1, k_2, ..., k_d$. This means that the eigenvalue $\mu_{\alpha_1 \alpha_2 ... \alpha_d}$ depends significantly on the order of its indices $\alpha_1$, $\alpha_2$, ..., $\alpha_d$. Then the degeneracy of the eigenvalue $\mu_{\alpha_1 \alpha_2 ... \alpha_d}$ is equal to $2^{d-r}$ where $r$ is the number of indices that are equal to one. As before, the same is true when we account for the nearest neighbors only.

Let us summarize. First, the degeneracy of the eigenvalues increases rapidly when the dimension of the lattice $d$ increases. Second, the long-range interaction does not remove the degeneracy that has place when we account for the interaction with the nearest neighbors only. Third, in the case of a multidimensional lattice an anisotropic interaction decreases the degeneracy by $d!$ times.

Concluding we would like to note a fact that is significant when $d \geq 2$. For simplicity, we suppose that the interaction is isotropic and we account for the nearest neighbors only. If $L$ was even, there would be a lot of zero-valued eigenvalues. For example, in the case of the two-dimensional system there are $L - 2$ pairs of the indices $\alpha$ and $\beta$, for which $\lambda_\alpha(1) = -\lambda_\beta(1)$ and $\mu_{\alpha\beta} = 0$. For the three-dimensional system there are about $L^2$ zero-valued eigenvalues. When $d$ increases the degeneracy of the zero-valued eigenvalue increases as $L^{d-1}$. Everywhere above we examined the odd values of $L$. In this case, the degeneracy of the zero-valued eigenvalue was not so large. However, when $L$ increases the number of the eigenvalues with close to zero values increases rapidly. This fact becomes evident when we turn to the case of the continuous spectrum.

In the asymptotic limit $L \to \infty$, it is more effective to examine not the discrete spectrum but the eigenvalue density $P(\mu)$. The analysis of Section 3 shows that in the most of the discussed cases the function $P(\mu)$ has singularities (the logarithmic and power divergences). We have shown that the long-range interaction leads to a significant increase of the degeneracy in some regions of the continuous eigenvalue spectrum.

## Acknowledgements

This work was supported by State Program of Scientific Research Institute for System Analysis, Russian Academy of Sciences, project no. 0065-2019-0003.